\documentclass[twocolumn,showpacs,preprintnumbers,superscriptaddress,amsmath,floatfix,amssymb,prl,nofootinbib]{revtex4}
\usepackage[colorlinks=true]{hyperref}
\usepackage{graphicx}
\newcommand{\comment}[1]{}

\newcommand{\Tr}{ {\rm Tr} \, }

\newcommand{\be}{\begin{equation}}
\newcommand{\ee}{\end{equation}}

\begin{document}
\sloppy

\title{Holonomy potential and confinement \\ from 
a simple model of  the gauge topology  }

\author{E. Shuryak}
\email{Edward.Shuryak@stonybrook.edu}
\affiliation{Department of Physics and Astronomy, Stony Brook University, Stony Brook 11794,USA}

\author{T. Sulejmanpasic}
\email{tin.sulejmanpasic@ur.de}
\affiliation{Institut f\"{u}r Theoretische Physik, Universit\"{a}t Regensburg, D-93040 Regensburg, Germany}

\begin{abstract}
We discuss an ensemble of topological solitons -- instanton-dyons and antidyons - in $SU(2)$ pure gauge theory at finite temperatures above and below the deconfinement phase transition temperature. The main focus is on the combined effect of this ensemble
on the so called effective holonomy potential, which drives the confinement/deconfinement phase transition. Using a simple 
model with excluded volume and lattice data on caloron density
we find that repulsive part of the potential is robust enough to induce the phase transition
at the right temperature.
Model's predictions -- the holonomy potential, electric and magnetic screening masses as a function of $T$ -- are in qualitative agreement 
with the
available lattice data. Further predictions are  densities of various dyon types as a function of temperature: while some lattice measurements of them had been made,
much more accurate data are needed to test these predictions.
\end{abstract}
\maketitle
\section{Introduction}
   The expectation value of the Polyakov loop $P=P\exp[ i \oint A_0d\tau]$ 
   is an order parameter of confinement/deconfinement phase transition in pure Yang Mill theory. 
    The gauge invariant integral over the
   thermal (Matsubara) circle is also called \emph{holonomy}.  Its average value
   and  fluctuations are studied numerically in lattice simulations, or   
   described by an effective potential   $U( P)$ fitted to lattice data, see e.g.  \cite{Pisarski:2001pe}. This potential plays a prominent role in current models of 
   finite temperature QCD, such as Polyakov-Nambu-Jona-Lasinio  model \cite{Fukushima:2003fw,Ratti:2005jh}, yet  the understanding of its origin is still missing. Some recent attempts to understand the structure of this potential have been made, both using the lattice gauge theory and using functional renormalisation group method \cite{Diakonov:2012dx,Greensite:2012dy, Haas:2013qwp}.
   
 \emph{Instantons},  the 4-dimensional solitons carrying the topological charge, are known to be important component of the gauge field
 theory. They have been used successfully in the description 
 of many nonperturbative phenomena  (for review see \cite{Schafer:1996wv}), including $U(1)$ and $SU(N_f)$ chiral symmetry breaking,
 but not confinement. At  nonzero asymptotic holonomy $A_0(r\rightarrow \infty)\ne 0$ the instanton solution is modified and as a result the instanton is split into $N_c$ constituents known
 as KvBLL self-dual instanton-dyons \cite{Lee:1997vp,Kraan:1998sn}. 
 
 Interactions and statistical ensemble of instanton-dyons and antidyons 
 have been studied qualitatively
    using neutral dyon-antidyon pairs in \cite{Shuryak:2012aa}, recently complemented by 
    first numerical simulations \cite{Faccioli:2013ja}. These works used holonomy as an input, defining masses and other
    properties of the instanton-dyons. They
   had studied the interaction between self-dual and anti--self-dual
   sectors  via  light fermions, and focused on the fermion collectivization  and chiral symmetry breaking phase transition.
In this paper we  turn our attention to the back reaction of the topology {\em on} the holonomy
potential, via a simple model with excluded volume, much in a spirit of  van der Waals theory of non-ideal gases.
 
  Two distinct sectors of instanton interaction are treated quite differently. Pure sefl-dual (or anti--self-dual) sector 
  has classicaly degenerate {\em moduli spaces} parameterized by collective coordinates:
   their studies are rather complete, and their
    nontrivial geometry  were extensively studied
  in mathematical literature since the 1970s by Atiyah, Hitchin and others. 
 
  The instanton--anti-instanton  ($\bar{I}I$) configurations   
  are studied significantly less. The
  moduli spaces are in this case substituted by the ``streamline configurations" \cite{Balitsky:1986qn,Verbaarschot:1991sq}
  which smoothly interpolate between the separated instanton--anti-instanton pair and the perturbative fields.
  Close $\bar{I}I$ pairs correspond to weak fields, which cannot be treated  semiclassically and should be subtracted from the semiclassical configurations.
  This  physical idea has been implemented in  the Instanton Liquid Model  
via an ``excluded volume", which generates a repulsive core and  stabilizes the density.  

In a few important cases, in which the partition function is independently known,
   such subtraction can be performed exactly, $without$  any parameters. 
The $\bar{I}I$  pair  contribution to the partition function in QM instanton problem has been done via the analytic continuation in the coupling constant $g^2\rightarrow -g^2$ by  Bogomolny \cite{Bogomolny:1980ur} and  Zinn-Justin \cite{ZinnJustin:1982td} (BZJ), who verified it via  known semiclassical series. Another
 analytic continuation has been used by Balitsky and Yung  \cite{Balitsky:1986qn} for  supersymmetric
  quantum mechanics. 
 
    Recently Poppitz, Sch\"afer and \"Unsal (PSU) \cite{Poppitz:2012nz,Poppitz:2012sw} used BZJ
   approach in the $N=1$ Super-Yang-Mills theory on $R^3\times S^1$,
   observing that the result obtained matches exactly the result derived  via supersymmetry \cite{Davies:2000nw}.
PSU papers are the most relevant for this work, as they focus on the  instanton-dyons (referred to as instanton-monopoles in their work).    The issue in this case is  more complicated
than for $\bar{I}I$:   dyon-antidyon pair always has uncompensated charges, magnetic or electric. Studies of the 
corresponding ``streamline configurations" is  yet to be done.

 In a carefully tuned weakly coupled, supersymmetric setting, softly broken by a small  gluino mass,
PSU \cite{Poppitz:2012nz,Poppitz:2012sw} have calculated the contribution of instanton-dyons and
 neutral instanton-dyon pairs. (They call  them {\em neutral bions}, but they are also known as
  $\bar{L}L, \bar{M}M$ molecules in other works.)  They observed that the repulsion in the latter term 
  can overcome the former one and drive
   the confinement/deconfinement phase transition.
 
 Supersymmetry forces the perturbative holonomy potential to be canceled, between contributions of gluons and (periodic) gluinos. 
 Pure Yang-Mills  (we are interested in) has no gluinos and there is no cancellations of the perturbative potential, which prefers the deconfined phase. The  QCD-like theories with thermally compactified fermion (i.e. fermions with anti-periodic temporal boundary condition) further add to this perturbative deconfining potential.     
In order to generate confinement, one needs to find some nonperturbative mechanism which is strong enough to compete with the perturbative holonomy potential.

 In this work we show  that  the free energy of the instanton-dyons can
 induce  holonomy potential which produces
 confinement in the pure gauge theory (using as an example the simplest SU(2) case)
 in qualitative agreement with the lattice data.  The central role is played by
  the effective repulsive interaction between dyon-antidyon, as in the  PSU works. We should stress, however, that PSU in \cite{Poppitz:2012sw} did consider non supersymetric, pure Yang-Mills case. We comment on the similarities and the differences of their work and ours later in the text.

\begin{figure}[t]
  \begin{center}
  \includegraphics[width=7cm]{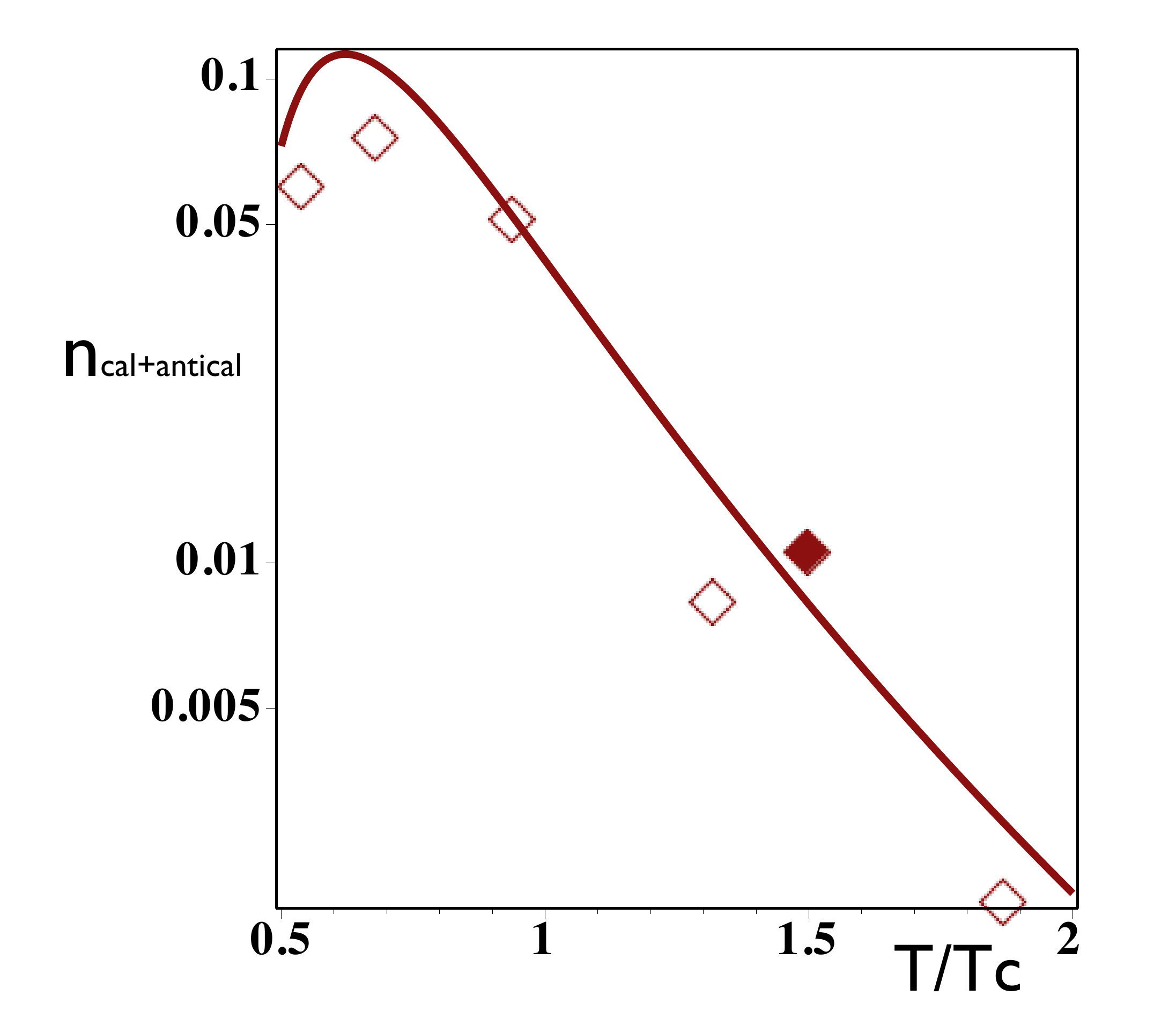}
  \caption{Caloron density as a function of $T/T_c$. The solid curve is the semiclassical fit $n_{\text{cal}}=K S_{cal}^2 e^{-S_{cal}}$ in units of $T$ with parameters $K=0.024$, $S_{cal}=8\pi^2/g^2(T)$, open (filled) points are the lattice data from \cite{Ilgenfritz:2006ju} (\cite{Bornyakov:2008im}). }
  \label{fig_cal}
  \end{center}
\end{figure}

For definiteness, we  start with fixing global parameters of the $SU(2)$  gauge theory topology, using
lattice works \cite{Ilgenfritz:2006ju} and \cite{Bornyakov:2008im}. The finite-$T$ instantons or  \emph{calorons}
possess only the topological charge -- they have neither 
  electric nor magnetic charges and thus no direct coupling to either electric or magnetic holonomies $b,\sigma$ (see next section). They
 can be identified on the lattice 
 via a number of well developed methods, e.g. the so called ``cooling".
In Fig.\ref{fig_cal} we present the data from \cite{Ilgenfritz:2006ju}  and their fit by the semiclassical expression for the dimensionless density 
(in units of $T^3$) of instantons/calorons 
\be n_{cal+\bar{cal}}= K S_{cal}^4 e^{-S_{cal}}, \,\, S_{cal}={22 \over 3} \ln \left({T\over \Lambda }\right)\ee
with parameters\footnote{Lattice practitioners usually fix ``physical units" via $T=0$ string tension, to which the QCD value is ascribed.
For pure gauge SU(2) one then finds $T_c\approx 270\, MeV$. } $K=0.024, \Lambda/T_c= .36$. The caloron action at $T_c$ is 7.50, 
so per dyon it makes $S_d=S_{cal}/2=3.75$, which
gives an idea how semiclassical the discussed objects are. (SU(3) instantons have actions $S_{cal}\approx 12$ or $S_d=S_{cal}/3\approx 4$, quite close in magnitude.) After those parameters are fixed, one knows semiclassical densities of the dyons and their pairs, as we explain in detail below.

\section{The holonomy potential}
In two-color $SU(2)$ gauge theory there is only one diagonal color matrix, thus one electric and one magnetic holonomy. An electric holonomy is closely related to the local Polyakov loop, which for SU(2) group can be parametrized by a single (space dependent) angle $\theta=v/T$
\be
\frac{1}{2}\Tr{P(x)}=\cos\left(\frac{v(x)}{2T}\right)\;,\quad  b={4\pi^2 \over g^2}\left({v\over \pi T} - 1\right)  
\ee
where $P(x)=\mathcal{P}\exp\left(i\oint A_0d\tau\right)$. The re-definition of $v$ into field $b$, by a shift and re-scaling, is done for further convenience (note that our $b$-field is labeled $b'$ in \cite{Poppitz:2012sw}).
Zero $v$ means trivial (deconfining) holonomy, zero $b$ means confinement.

Magnetic potential (or magnetic holonomy) $\sigma$ was introduced by Polyakov Ref.\cite{Polyakov:1976fu}, together with an observation that
\begin{multline} \exp\left(-{e_1e_2 \over 4\pi|\vec x - \vec y|}\right)=\\=\int D\sigma\; \exp\left[- (\vec\partial \sigma)^2/2+ ie_1\sigma(\vec x)+ie_2\sigma(\vec y)\right]
\end{multline}
which expresses Coulomb forces via corresponding potential field $\sigma$. Since our objects have both electric
 and magnetic charges and potentials, the trick is used twice. The amplitudes for the 4 types of dyons take the exponental form \cite{Poppitz:2012sw}

\begin{align}
&\mathcal{M}\sim e^{-b+i\sigma-S_d}\;,&&\bar{\mathcal{M}}\sim e^{-b-i\sigma-S_d}\\
&\mathcal{L}\sim e^{b-i\sigma-S_d}\;,&&\bar{\mathcal{L}}\sim e^{b+i\sigma-S_d}\;.
\end{align}
Note that there is no $i$ in the electric part. This behavior is \emph{not} however because of the Euclidean formulation of the theory, as the real charge couples the electric potential $A_0$ as $\exp(i\int A_0d\tau)$ both in Euclidean and Minkowski space-time (see the discussion in \cite{Poppitz:2012sw}). In fact the coupling to $b$-field is through the action of instanton-dyons which is $b$ dependent, and it always appears in the combination $S_d\pm b$ which is the action of the $L$ and $M$ dyon. 
Contributions of all four instanton-dyons can be combined into the following effective potential
\be
V_{eff}=-4n_d^0\cos(\sigma)\cosh(b)
\ee
where we use a ``generic instanton-dyon density"  $n^0_d=(n_{cal})^{1/N_c}$ as a prefactor. The Coulomb gas partition function is recovered if $b,\sigma$ are 
integrated out with a kinetic energy $\mathcal{L}_k=\frac{g^2T}{2(4\pi)^2} \left[(\partial_i\sigma)^2+(\partial_ib)^2\right]$.

The next step is an introduction of the binary repulsion and the excluded volume in $\bar{L}L$ and $ \bar{M}M$ channels. Subtracting pair contraction of instanton-dyon--antidyon the partition function takes the form
\begin{multline}\label{eq:sum}
\sum_{k,N,\bar N}\frac{1}{k! N!\bar N!}\left(-\int d^3x\;e^{\pm2b(x)} n_dn_d \mathcal{A}\right)^k \\\left(\int d^3x\;e^{\pm b\pm i\sigma}n_d\right)^N\left(\int d^3x\;e^{\pm b\mp i\sigma}n_d\right)^{\bar N}
\end{multline}
where $\mathcal A$ is the subtracted amplitude
\be\label{eq:amp}
\mathcal A=4\pi\int_{0}^{r_0}dr\;r^2e^{-V_{\bar{d}d}( r)}
\ee
where the upper sign refers to $M$ type and the lower sign to $L$ type instanton-dyons  and anti-dyons and.
The effective dyon-antidyon interaction is Coulombic at large distances $V_{\bar{d}d}=-8\pi \beta/g^2r$ and regulated at small distances by some particular scheme, e.g. $1/r\rightarrow 1/\sqrt{r^2+a^2}$ needed to stabilise the coulomb plasma.
If the objects were for example atoms this regularization would suffice. It is  the instantonic nature of these objects
-- not regularisation -- which requires the additional subtraction as the close pair of instanton-dyon and antidyon are perturbative fields, and it is this subtraction which brings  in a
new parameter $r_0$ or $\mathcal A$.

The effective potential then becomes
 \be\label{eq:naivepotential}
V_{eff}= 2n_d\left(-2\cos\sigma\cosh b+n_d\mathcal{A}\cosh(2b)\right)\;.
\ee
and has a minimum at
\be
\sigma_{min}=0,b_{min}=\left\{\begin{array}{ll}
0\; &\text{for $2n_d\mathcal A\ge 1$}\\
\cosh^{-1}\left(\frac{1}{2n_d\mathcal A}\right)\;&\text{for $2n_d\mathcal A<1$}
\end{array}\right.
\ee
This is the key point: confining regime (the one with $b_{min}=0$) can only be reached if 
 the instanton-dyon density is sufficiently large.

The third step is inclusion of  the one-loop perturbative potential \cite{Gross:1980br}
\be
V_{pert}=\frac{\pi^2}{12}\left(1-\frac{b^2}{S_d^2}\right)^2
\ee
which by itself prefers a trivial holonomy $v=0,b=S_d$. Including this terms  one finds that
the phase transition is  pushed to higher densities. By demanding that the second order coefficient of $b$ vanishes, we obtain the equation for critical temperature of confinement/deconfinement transition 
\be
-2 n_d^c+4 \mathcal{A} (n_d^c)^2 - \frac{\pi^2 }{6 S_d^2}=0,
\ee
Note that all (dimensionfull) quantities above are in units of $T$. Since we take that $n_d^c=\sqrt{n_{cal}(T_c)}=\sqrt{K/2}S_{cal}(T_c)^2e^{-S_{cal}(T_c)}\approx 0.145$ we need to have $\mathcal A\approx 5$.  While  $\mathcal A$ is much larger than the dyon volume $\sim 4\pi/3(1/\pi)^3$, including the  Coulomb enhancement factor $\sim \exp(O(S_d))$ expected for the pair of ``half-annihilated" dyon-antidyon  gives this parameter the correct order of magnitude.

\begin{figure}[t]

  \includegraphics[width=7cm]{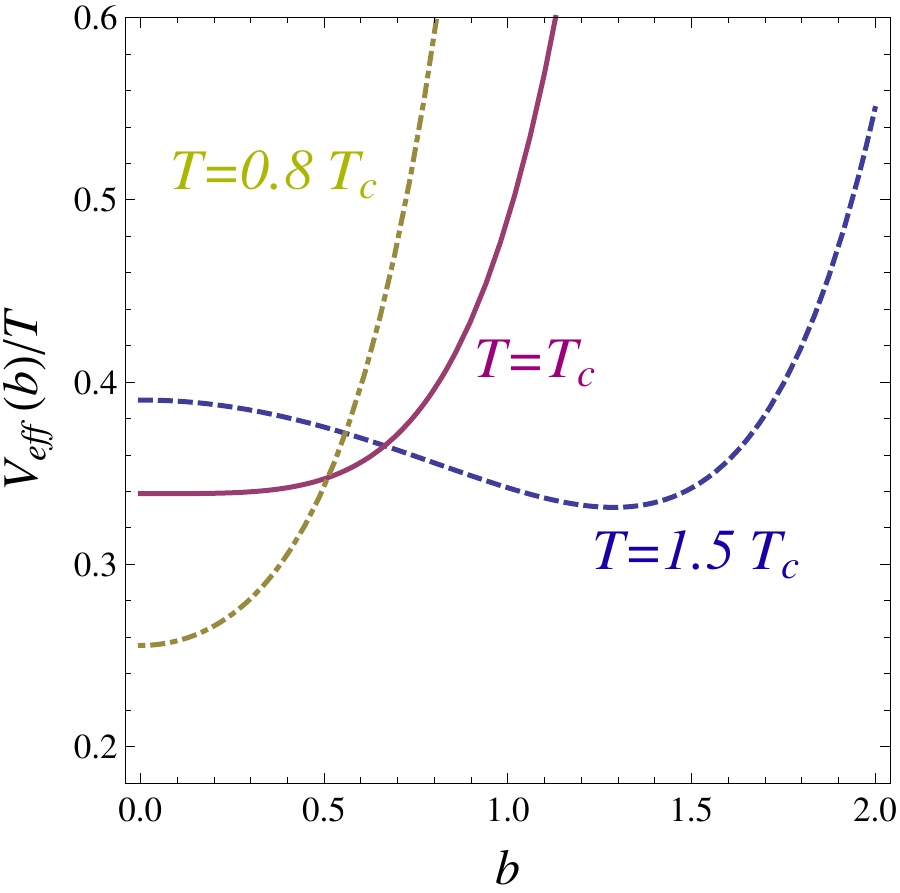}

   \vspace{.5cm}
   
      \hspace{.5cm}\includegraphics[width=6.5cm]{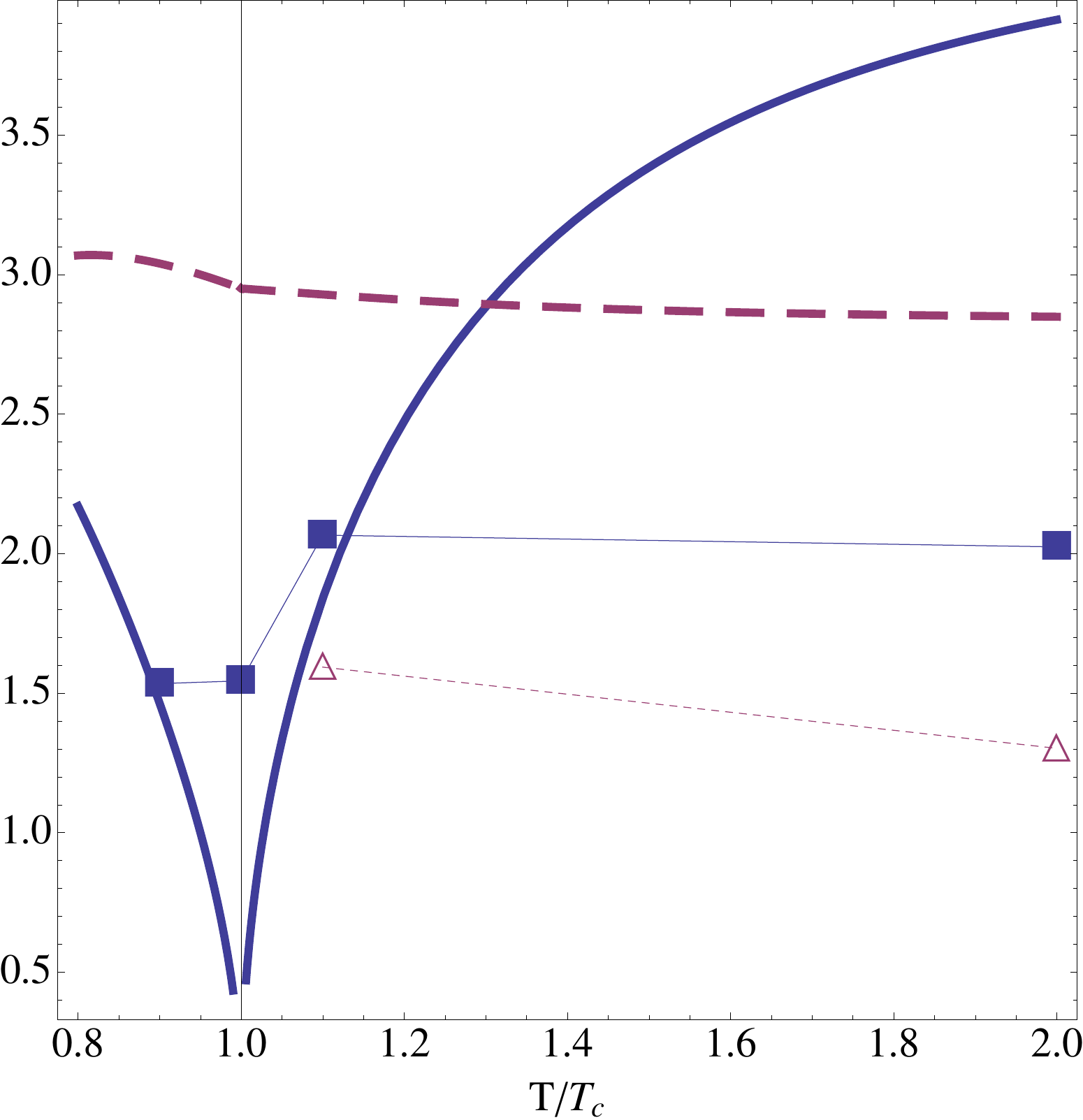}

  \caption{The upper plot shows 
  the effective potential $V_{eft}(b)/T$ \eqref{eq:potential} for $T/T_c=0.8,1,1.5$ shown by the dashed,solid and dot-dashed lines, respectively. The plot shows electric $m_E/T$ and magnetic $m_M/T$ screening masses versus temperature, indicated by the solid and 
  dashed lines, respectively. 
  Thick lines are our model, the data points are from lattice propagators \cite{Bornyakov:2010nc}, the lines connecting data points are shown simply for their identification.  }
  \label{fig_pot}

\end{figure}

Let us improve the model a bit by noting that the excluded volume 
 should be different for $M$ and $L$ dyons, and in fact should depend on the holonomy $b$, in such a way that $r_0\propto (1\pm b/S_d)^{-1}$. To account for this $\mathcal A (1\pm \frac{b}{S_d})^{-3}$, where $\pm$ refers to the instanton-dyon with factors $e^{\mp b}$ respectively. Further each dyon carries with it a moduli space metric parameter in front of the densities, which should also be included and that can be obtained by replacing $n_de^{\mp b}\rightarrow n_d(1\pm \frac{b}{S_d})e^{\mp b}$. This results in a somewhat longer expression for the effective potential 
\begin{multline}\label{eq:potential}
V_{eff}=-2n_d\left[\left(1+\frac{b}{S_d}\right)e^{-b}+\left(1-\frac{b}{S_d}\right)e^{b}\right]\cos\sigma\\+n_d^2\mathcal{A}\left[\frac{e^{2b}}{1-\frac{b}{S_d}}+\frac{e^{-2b}}{1+\frac{b}{S_d}}\right]+\frac{\pi^2}{12}\left(1-\frac{b^2}{S_d^2}\right)\;.
\end{multline}
The effect of these factors helps the phase transition to occur and similar analysis as before gives $\mathcal A\approx 2.3$. The resulting holonomy potential at $T/T_c=0.8,1.,1.5 $ are shown in the upper plot of Fig. \ref{fig_pot}.

To conclude this section, let us mention that the potentials \eqref{eq:naivepotential} and \eqref{eq:potential} are similar to the one discussed in \cite{Poppitz:2012sw} for the case of pure Yang-Mills. We briefly comment on two important differences: the ``excluded volume''  parameter $\mathcal A$ and the absence of the magnetic bion $\cos(2\sigma)$. The magnetic bion term in our analysis can be obtained in a similar way to the subtraction, by a virial-type of expansion. However since the $M\bar L$ and $L\bar M$ pair repel, this term will be accompanied by a coulomb suppression term, rather then coulomb enhancement. Therefore we believe that such terms in the effective action can be safely neglected. We also note that the coefficient $\mathcal A$ should formally be temperature dependent, at least through the coupling constant. The naive way to incorporate this dependence would seem to suggest that this term is growing with temperature, which is unphysical, as it is expected that instanton-dyon effects and charges become screened and the effects on the holonomy become less and less important. Since our paper focuses  at describing the physics around $T_c$, we assumed the parameter $\mathcal A$ to be constant.

\section{Predictions}
Now that the model is fully defined,  some of its predictions can be compared to the lattice data. We start with
two important observables usually not associated with topology and measured directly from
propagators:  the (mean-field, bare) electric and magnetic screening masses
\begin{align} 
&m_E^2=({4S_d}){\partial^2 V_{eff}\over \partial b^2}\Bigg|_{\substack{b=b_{min}\\\sigma=\sigma_{min}}}\;\\
&m_M^2=({4S_d})^2{\partial^2 V_{eff}\over \partial \sigma^2}\Bigg|_{\substack{b=b_{min}\\\sigma=\sigma_{min}}}
\end{align}
where $b_{min},\sigma_{min}$ are values where the potential takes its minimum value. 

Their comparison with lattice data is shown in the lower plot of Fig. \ref{fig_pot}.
Our model produces phase transition  of the second order, as is known to be the case for SU(2) theory. This implies that the bare electric mass  must
vanish
at $T_c$. While this is indeed evident from holonomy distributions observed on the lattice (see e.g. Fig. 4 of \cite{Ilgenfritz:2006ju}),
the electric mass obtained from gluon propagators \cite{Bornyakov:2010nc} (shown by the open squares in Fig. \ref{fig_pot}) indicate only partial downward shift. Note, however, that the lattice measurements are for the full, physical mass of the gluonic propagator, and that our model has infinitely many $b$-field self-couplings which should renormalize the mass especially around $T_c$ where the fluctuations are most important (and create critical indices
different from mean field ones, as e.g. observed on the lattice in \cite{Hubner:2008ef}).

The behavior of the magnetic mass in the model has a fairly smooth behavior through the phase transition, with a small kink at $T_c$ which is due to the dependence of the magnetic mass on the density the $M$ and $L$ dyons which exhibits a non smooth behavior due to the 2nd order phase transition. It can be seen that the SU(2) data on magnetic mass  \cite{Bornyakov:2010nc}, shown by triangles, have only  two points at
two temperatures.  However analogous but much more complete SU(3) data shows a smooth 
$m_M(T)$ behavior, in spite of the 1-sf order transition in this theory. 
Note that the absolute scale of the masses are about factor 2 off, which is hardly surprising since the  normalization was done
in a topological sector with similar uncertainty (see Fig.1).  The ratio $m_M/m_E$ predicted by the model and from the lattice
agree rather well. We thus conclude that the model performs at a qualitative level well enough. 

More specific  predictions of the model are of course the dimensionless (i.e. in units of $T$) densities of monopole-dyons of  particular types:
\begin{multline} n_{M,L}=  n_d (1\pm b_{min}/S_d) e^{\pm b_{min}} \\
\times\left( 1-{1 \over 2}\mathcal{A}  n_d (1\pm b_{min}/S_d)^{-2} e^{\pm b_{min}}\right)
\end{multline}

\begin{figure}[t]
  \begin{center}
  \includegraphics[width=7cm]{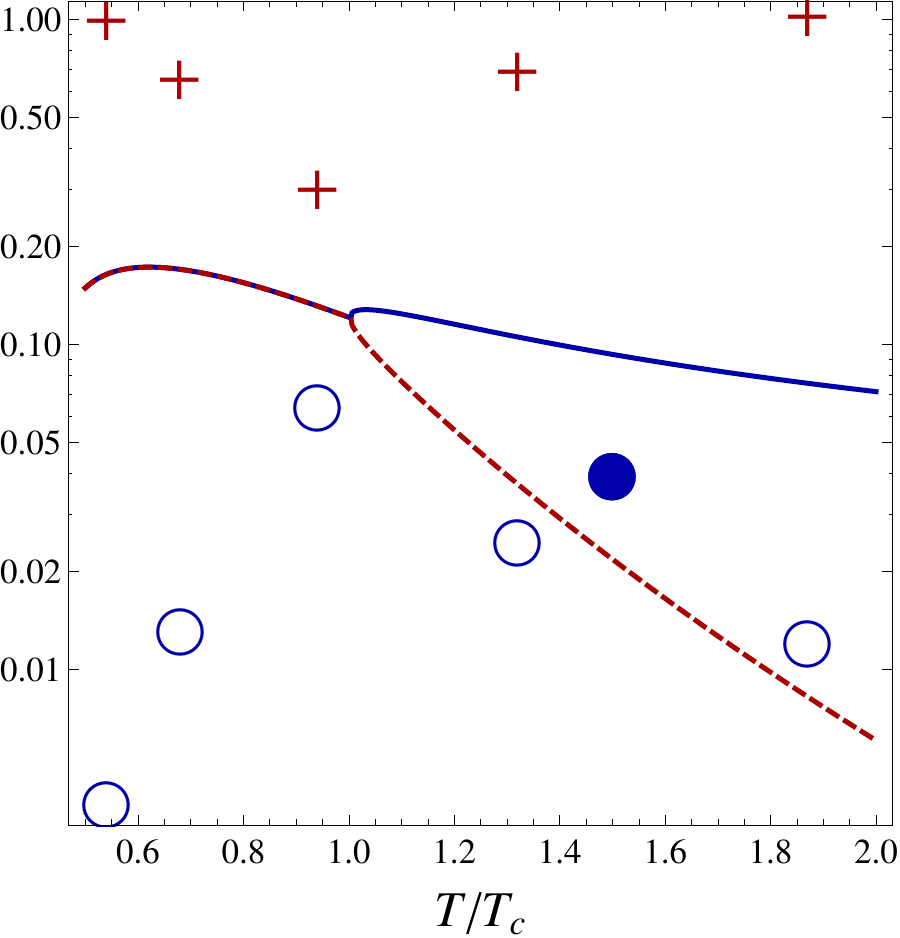}
  \caption{Prediction of the model for the temperature dependence of the density of the instanton-dyons are shown by the lines, those with
  solid and dashed lines are for $M,L$ type dyons, respectively. Open (filled) circles show identified $M$-type dyons from ref.
 \cite{Ilgenfritz:2006ju} (\cite{Bornyakov:2008im}). The crosses show ``unidentified topological objects" from   \cite{Ilgenfritz:2006ju}. Circles and crosses provide the lower and the upper bound for the dyon density. }
  \label{fig_top}
  \end{center}
\end{figure}

The densities are shown as curves in Fig. \ref{fig_top}.  At $T>T_c$ there are two curves, identifying lighter (and thus more numerous) $M$
dyons as well as $L$ ones. Below $T_c$ the curves collapse into one single curve as the potential develops a minimum at $b=0$. 

Two examples
of the best lattice  efforts to determine the densities in  SU(2) pure Yang-Mills theory
are refs. \cite{Ilgenfritz:2006ju} and \cite{Bornyakov:2008im}. They use  $24^3\times6$ and $20^3\times 4$ lattices, the former with couplings tuned to three points with $T<T_c$ and two $T>T_c$. The $L$ and $M$ dyons can in principle be identified by their electric and magnetic charges as well as
by the value of the Polyakov loop at their centers taking values $\pm 1$ for the two type of dyons. In  practice, the heavy $L$ dyon has such a small size that
it cannot be identified by this method at current lattices. The $M$ instanton-dyons, however, were identified and the results are shown as circles in Fig. \ref{fig_top}. The authors of \cite{Ilgenfritz:2006ju}
 note however that the efficiency of $M$-dyon identification
is quite low, and  depends on the temperature. Crosses
 show ``unidentified topological objects" from the same work  \cite{Ilgenfritz:2006ju}, to be taken as an upper bound.  As one can see from the plot,  our predictions lie between the lower and upper bound. More work is obviously needed to test these predictions. 
  Apart from the densities of the instanton-dyons, quite
 valuable would be lattice studies of their  spatial correlations, as those can
 help understand their interactions as well.

Concluding this work, let us comment that while the model proposed is rather schematic, it represents a potentially  important link between
the gauge topological sector -- instanton-dyons -- and confinement in QCD-like theories. 
Its more quantitative forms and
generalization to QCD with fermions of different kind seem to be straightforward. Adding fundamental fermions would induce extra binding of
$\bar{L}L$ pairs via fermonic zero modes: this will shift the deconfinement transition to lower $T$ and/or higher instanton-dyon densities, as is indeed
observed phenomenologically. We intend to study topology and all related phenomena quantitatively, by a direct Monte-Carlo simulations, elsewhere.  

{\bf Acknowledgements}
TS would like to thank E. Poppitz on the discussions and insights as well as hospitality during the recent visit to the University of Toronto. ES is supported by the U.S. Department of Energy Grant DE-FG-88ER40388. TS is supported by the BayEFG Scholarship.


\begin{thebibliography}{20}

\bibitem{Pisarski:2001pe} 
  R.~D.~Pisarski,
  Nucl.\ Phys.\ A {\bf 702}, 151 (2002)
  [hep-ph/0112037].

\bibitem{Fukushima:2003fw} 
  K.~Fukushima,
  Phys.\ Lett.\ B {\bf 591}, 277 (2004)
  [hep-ph/0310121].
\bibitem{Ratti:2005jh} 
  C.~Ratti, M.~A.~Thaler, W.~Weise ,
  Phys.\ Rev.\ D {\bf 73}, 014019 (2006)
  [hep-ph/0506234].
  
\bibitem{Diakonov:2012dx} 
  D.~Diakonov, C.~Gattringer and H.~-P.~Schadler,
  JHEP {\bf 1208}, 128 (2012)
  [arXiv:1205.4768 [hep-lat]].
  
\bibitem{Greensite:2012dy} 
  J.~Greensite,
  Phys.\ Rev.\ D {\bf 86}, 114507 (2012)
  [arXiv:1209.5697 [hep-lat]].
 
 
\bibitem{Haas:2013qwp} 
  L.~M.~Haas, R.~Stiele, J.~Braun, J.~M.~Pawlowski and J.~Schaffner-Bielich,
  PhysRevD.87.076004,2013
  [arXiv:1302.1993 [hep-ph]].
  
  
\bibitem{Schafer:1996wv} 
  T.~Schafer, E.~V.~Shuryak ,
  Rev.\ Mod.\ Phys.\  {\bf 70}, 323 (1998)
  [hep-ph/9610451].

  
\bibitem{Lee:1997vp}
  K.~M.~Lee and P.~Yi,
  Phys.\ Rev.\  D {\bf 56}, 3711 (1997)
  [arXiv:hep-th/9702107].

\bibitem{Kraan:1998sn}
  T.~C.~Kraan and P.~van Baal,
  Phys.\ Lett.\  B {\bf 435}, 389 (1998)
  [arXiv:hep-th/9806034].



\bibitem{Shuryak:2012aa} 
  E.~Shuryak, T.~Sulejmanpasic ,
  Phys.\ Rev.\ D {\bf 86}, 036001 (2012)
  [arXiv:1201.5624 [hep-ph]].
  
\bibitem{Faccioli:2013ja} 
  P.~Faccioli, E.~Shuryak ,
  arXiv:1301.2523 [hep-ph].
  
\bibitem{Verbaarschot:1991sq} 
  J.~J.~M.~Verbaarschot,
  Nucl.\ Phys.\ B {\bf 362}, 33 (1991)
  [Erratum-ibid.\ B {\bf 386}, 236 (1992)].
  
\bibitem{Balitsky:1986qn} 
  I.~I.~Balitsky, A.~V.~Yung ,
  Phys.\ Lett.\ B {\bf 168}, 113 (1986).


\bibitem{Bogomolny:1980ur} 
  E.~B.~Bogomolny,
  Phys.\ Lett.\ B {\bf 91}, 431 (1980).
  
\bibitem{ZinnJustin:1982td} 
  J.~Zinn-Justin,
  Nucl.\ Phys.\ B {\bf 218}, 333 (1983).

\bibitem{Poppitz:2012sw} 
  E.~Poppitz, T.~SchŠfer and M.~Unsal,
  JHEP {\bf 1210}, 115 (2012)
  [arXiv:1205.0290 [hep-th]].

\bibitem{Poppitz:2012nz} 
  E.~Poppitz, T.~SchŠfer and M.~Unsal,
  arXiv:1212.1238 [hep-th].


 
  
\bibitem{Davies:2000nw} 
  N.~M.~Davies, T.~J.~Hollowood, V.~V.~Khoze ,
  J.\ Math.\ Phys.\  {\bf 44}, 3640 (2003)
  [hep-th/0006011].

\bibitem{Ilgenfritz:2006ju} 
  E.~-M.~Ilgenfritz, B.~V.~Martemyanov, M.~Muller-Preussker and A.~I.~Veselov,
  Phys.\ Rev.\ D {\bf 73}, 094509 (2006)
  [hep-lat/0602002].

\bibitem{Bornyakov:2008im} 
  V.~G.~Bornyakov, E.~-M.~Ilgenfritz, B.~V.~Martemyanov and M.~Muller-Preussker,
  Phys.\ Rev.\ D {\bf 79}, 034506 (2009)
  [arXiv:0809.2142 [hep-lat]].


\bibitem{Polyakov:1976fu} 
  A.~M.~Polyakov,
  Nucl.\ Phys.\ B {\bf 120}, 429 (1977).
  
  

\bibitem{Gross:1980br} 
  D.~J.~Gross, R.~D.~Pisarski, L.~G.~Yaffe ,
  Rev.\ Mod.\ Phys.\  {\bf 53}, 43 (1981).




\bibitem{ZinnJustin:1981dx} 
  J.~Zinn-Justin,
  Nucl.\ Phys.\ B {\bf 192}, 125 (1981).



\bibitem{ZinnJustin:1981jn} 
  J.~Zinn-Justin,
  In *Sant Feliu De Guixols 1981, Proceedings, Non-perturbative Aspects Of Quantum Field Theory*, 179-217
  
\bibitem{Balitsky:1985in} 
  I.~I.~Balitsky, A.~V.~Yung ,
  Nucl.\ Phys.\ B {\bf 274}, 475 (1986).



  
  
  
  


\bibitem{Bornyakov:2010nc} 
  V.~G.~Bornyakov, V.~K.~Mitrjushkin and ,
  Phys.\ Rev.\ D {\bf 84}, 094503 (2011)
  [arXiv:1011.4790 [hep-lat]].

\bibitem{Hubner:2008ef} 
  K.~Huebner, C.~Pica ,
  PoS LATTICE {\bf 2008}, 197 (2008)
  [arXiv:0809.3933 [hep-lat]].

\end{thebibliography}

\end{document}